%% file: Locally_recorded_decomp-arXiv-submission.tex
\documentclass[letterpaper,pra,floatfix,twocolumn,aps,10pt]{revtex4-1}
\input{riedelheader.tex}
\pdfoutput=1

\renewcommand{\draftmode}{0} 

\begin{document}



\title{Local Records and Global Entanglement: A Unique Multi-Partite Generalization of the Schmidt Decomposition}
\date{\today}
\author{C.~Jess~Riedel}
\affiliation{IBM Watson Research Center, Yorktown Heights, NY 10598}


\begin{abstract}
We show that there is a unique maximal decomposition of a pure multi-partite ($N>2$) quantum state into a sum of states which are ``locally orthogonal'' in the sense that the local reduced state for a term in the sum lives in its own orthogonal subspace for each subsystem.  Observers can make local measurements on any subsystem and determine which ``branch'' they are on.    The Shannon entropy of the resulting branch weights defines a new measure of global, GHZ-like entanglement, which is insensitive to local pairwise entangling operations and vanishes when there is no piece of information recorded at every subsystem.  In the bi-partite ($N=2$) case, this decomposition reduces to the (not necessarily unique) Schmidt decomposition and the entropy reduces to the entropy of entanglement.
\end{abstract} 

\maketitle


Suppose we are given a normalized quantum state $\vert \psi \rangle$ from a Hilbert space $\mathcal{H}$ composed of $N$ smaller finite-dimensional subsystems $\mathcal{H}^{(n)}$ that are tensored together:
\begin{align}
\label{eq:tensor-structure}
\vert \psi \rangle \in \mathcal{H} = \mathcal{H}^{(1)} \otimes \cdots \otimes \mathcal{H}^{(N)}.
\end{align}
We seek a preferred decomposition 
\begin{align}
\label{eq:decomp}
\vert \psi \rangle = \sum_i \sqrt{p_i} \vert \psi_i \rangle
\end{align}
of $\vert \psi \rangle$ expressed as a sum (weighted by the strictly positive values $\sqrt{p_i}$) of orthonormal vectors $\vert \psi_i \rangle$ such that the $\vert \psi_i \rangle$ are pairwise orthogonal on each subsystem.  More precisely, call a decomposition \emph{locally orthogonal} (LO) with respect to the tensor structure \eqref{eq:tensor-structure} if it satisfies
\begin{align}
\label{eq:locally-ortho}
\vert \psi_i \rangle \in \mathcal{H}^{(1)}_i \otimes \cdots \otimes \mathcal{H}^{(N)}_i.
\end{align}
where
\begin{align}
\label{eq:subspaces}
\mathcal{H}^{(n)} = \bigoplus_i \mathcal{H}^{(n)}_i 
\end{align}
is some decomposition of each $\mathcal{H}^{(n)}$ into orthogonal subspaces. (That is, $\braket{\mu}{\nu} = 0$ whenever $\ket{\mu} \in \mathcal{H}^{(n)}_i$, $\ket{\nu} \in \mathcal{H}^{(n)}_j$, and $i \neq j$.) To be suggestive, we call the $\ket{\psi_i}$ the \emph{branches} of the decomposition.

This decomposition is interesting because it means that the supports of the $i$-conditional local density matrices of any subsystem $n$,
\begin{align}
\rho^{(n)}_i \equiv \mathrm{Tr}_{\overline{n}} \big[ \vert \psi_i \rangle \langle \psi_i \vert \big]
\end{align}
are contained in the orthogonal subspaces $\mathcal{H}^{(n)}_i$. (Here, the trace is over all subsystems except $n$.) Therefore, two observers can each make local measurements on any of the systems and each independently determine the branch.  

Now, note that any non-trivial LO decomposition can be coarse-grained to a new decomposition by simply combining branches through addition:
\begin{align}
\label{eq:coarse-graining}
\sqrt{p_{i'}}\ket{\psi_{i'}}, \sqrt{p_{i''}}\ket{\psi_{i''}}  \to \sqrt{p_{i}}\ket{\psi_{i}} = \big(\sqrt{p_{i'}}\ket{\psi_{i'}} + \sqrt{p_{i''}}\ket{\psi_{i''}} \big).
\end{align}
The new branch is contained in the coarse-grained subspace $\mathcal{H}^{(n)}_i = \mathcal{H}^{(n)}_{i'} \oplus \mathcal{H}^{(n)}_{i''}$, so the new decomposition is also LO.  
The main result of this article is that, for $N>2$, there is a unique maximum decomposition in the sense that all other LO decompositions are coarse-grainings of the maximal one.  In the case $N=2$, there may be multiple distinct LO decompositions which cannot be further fine-grained.  These are precisely the Schmidt decompositions, and the possibility of non-uniqueness (which is always associated with a degeneracy in the spectrum of the local density matrix) is a peculiar feature of $N=2$. 

The rest of this article is laid out as follows.  First, we define some preliminary objects.  Second, we prove uniqueness of the maximum LO decomposition by demonstrating a common fine-graining of any two given LO decompositions.  Third, we explicitly construct the maximum in the non-degenerate case.  Fourth, we will introduce a corresponding notion of global entanglement defined by the maximum.  Finally, we look at some examples.

\section{Preliminaries} \label{sec:preliminaries}

For a given decomposition $\ket{\psi} = \sum_i \sqrt{p_i} \ket{\psi_i} $, define the orthogonal projectors $Q^{(n)}_i$ to project onto the respective supports of $\rho^{(n)}_i$.  In other words,
\begin{align}
Q^{(n)}_i = I^{\overline{(n)}} \otimes \sum_{\mu \in T_i^{(n)}} \ket{\mu}_n \bra{\mu}  
\end{align}
which acts trivially otherwise except on subsystem $n$.  Here, $\mu$ indexes the local Schmidt basis of the branch $\ket{\psi_i}$ on subsystem $n$, and $T_i^{(n)}$ is the set of index values.  (When the spectrum of $\rho^{(n)}_i$ is degenerate there will be multiple choices of basis, but $Q^{(n)}_i$ is still unambiguous.) This means $\left[Q^{(n)}_i, Q^{(m)}_{j} \right] = 0$ for all $n$, $m$, $i$, and $j$.  

The condition for the decomposition to be LO is equivalent to
\begin{align}
Q^{(n)}_i Q^{(m)}_j \ket{\psi} = \delta_{i j} \sqrt{p_i} \ket{\psi_i}
\end{align}
for any $n$ and $m$.  We then define $Q_i \equiv Q^{(1)}_i \cdots \, Q^{(N)}_i$ so that $\ket{\psi_i} \in \Glob^{(1)}_i \otimes \cdots \otimes \Glob^{(N)}_i$ and
\begin{align}
\sum_i Q_i \ket{\psi} = \sum_i \sqrt{p_i} \ket{\psi_i} = \ket{\psi}.
\end{align}

The relation of coarse-graining forms a partial order on the set of decompositions, and a decomposition is called \emph{maximal} if it has no fine-graining.  In general, a partially ordered set may have multiple maximal elements, but when there is only one it is called a \emph{maximum}.
\section{Uniqueness} \label{sec:uniqueness}

Suppose $N>2$ and we are given two LO decompositions $\ket{\psi} = \sum_i \sqrt{p_i} \ket{\psi_i}$ and $\ket{\psi} = \sum_k \sqrt{\tilde{p}_k} \ket{\tilde{\psi}_k}$ with local subspaces $\Glob^{(n)}_i$ and $\tilde{\Glob}^{(n)}_k$.  Define the projectors $\tilde{Q}^{(n)}_k$ and $\tilde{Q}_k$ analogously to $Q^{(n)}_k$ and $Q_k$, and then consider that
\begin{align}\begin{split}
\label{eq:cross-corr}
\sqrt{\tilde{p_k}} Q^{(n)}_{i}  Q^{(m)}_{j} \ket{\tilde{\psi}_k} &= Q^{(n)}_{i}  Q^{(m)}_{j} \tilde{Q}^{(r)}_{k} \ket{\psi} \\
&= \tilde{Q}^{(r)}_{k} Q^{(n)}_{i}  Q^{(m)}_{j}  \ket{\psi} \\
&= \delta_{i j} \sqrt{p_i} \tilde{Q}^{(r)}_{k} \ket{\psi_i} 
\end{split}\end{align}
for any $r \neq n,m$.  Here it is crucial that $N>2$ to guarantee the existence of at least one such an $r$.

Since \eqref{eq:cross-corr} vanishes when $i \neq j$, regardless of $n$ and $m$, we know that 
\begin{align}
\ket{\tilde{\psi}_k} \in \bigoplus_i \left( \Glob^{(1)}_i \otimes \cdots \otimes \Glob^{(N)}_i \right).
\end{align}
(Otherwise $\ket{\tilde{\psi}_k}$ would have a non-zero component in $\Glob_{i_1}^{(1)} \otimes \cdots \otimes \Glob_{i_N}^{(N)}$ for some choice of $i_n$'s that are not all equal.  This would conflict with \eqref{eq:cross-corr}.) Therefore
\begin{align}\begin{split}
\label{eq:commute}
\sqrt{\tilde{p_k}} Q_i \ket{\tilde{\psi}_k} &= Q^{(n)}_i \sqrt{\tilde{p_k}} \ket{\tilde{\psi}_k} \\
&= Q^{(n)}_i \tilde{Q}^{(r)}_k \ket{\psi} \\
&= \tilde{Q}^{(r)}_k Q^{(n)}_i \ket{\psi} \\
&= \sqrt{p_i}  \tilde{Q}_k \ket{\psi_i} \\
\end{split}\end{align}
where the last line follows from reproducing the first three lines with the two decompositions exchanged.
We can then introduce a third decomposition 
\begin{align}
\label{eq:common-fine}
\ket{\psi} = \sum_{(k,i)\in R} \sqrt{\hat{p_{k,i}}} \ket{\hat{\psi}_{k,i}}
\end{align}
(which we will show to be a common fine-graining) where 
\begin{align}
\label{eq:joint}
\sqrt{\hat{p_{k,i}}} \ket{\hat{\psi}_{k,i}} \equiv \sqrt{\tilde{p_k}} Q_i \ket{\tilde{\psi}_k} = \sqrt{p_i} \tilde{Q}_k \ket{\psi_i}
\end{align}
and $R$ is defined to be the set of pairs $(k,i)$ for which \eqref{eq:joint} does not vanish.  Letting $\hat{Q}_{k,i}^{(n)} \equiv \tilde{Q}_{k}^{(n)} Q_{i}^{(n)}$, we confirm that this is a LO decomposition: 
\begin{align}\begin{split}
\hat{Q}_{k,i}^{(n)} \hat{Q}_{l,j}^{(m)} \ket{\psi} & =
\tilde{Q}_{k}^{(n)} Q_{i}^{(n)} \tilde{Q}_{l}^{(m)} Q_{j}^{(m)} \ket{\psi} \\
& = \tilde{Q}_{k}^{(n)} \tilde{Q}_{l}^{(m)} \delta_{i,j} \sqrt{p_i} \ket{\psi_i} \\
& = \tilde{Q}_{l}^{(m)} Q_{j}^{(m)} \tilde{Q}_{k}^{(n)} Q_{i}^{(n)}  \ket{\psi} \\
& = Q_{i}^{(n)} Q_{j}^{(m)} \delta_{k,l} \sqrt{\tilde{p_k}} \ket{\tilde{\psi}_i},
\end{split}\end{align}
which vanishes for all $n,m$ unless $(k,i) = (l,j)$. (Above we have made use of the fact, by \eqref{eq:commute}, that $Q_{i}^{(r)}$ and $\tilde{Q}_{k}^{(r)}$ commute when acting on $\ket{\psi}$ for any $r$.) In other words, the $\ket{\hat{\psi}_{k,i}}$ live in the refined subspaces $\hat{\Glob}_{k,i} \equiv \tilde{\Glob}_k \cap \Glob_i$.  Summing \eqref{eq:joint} over $k$ and $i$ confirms that this decomposition is a common fine-graining:
\begin{align}
\ket{\tilde{\psi}_k} = \sum_i \ket{\hat{\psi}_{k,i}} \quad , \quad \ket{\psi_i} = \sum_k \ket{\hat{\psi}_{k,i}}.
\end{align}

It's clear that any non-trivial common fine-graining \eqref{eq:common-fine} of two LO decompositions that are not identical must have strictly more non-zero branches than either of the two.  Combined with the fact that no LO decomposition may have more branches than the dimension of the smallest subsystem, this shows that there is a unique maximal LO decomposition.

Finally, we note when $N=2$ a Schmidt decomposition of $\ket{\psi}$ is an LO decomposition and is maximal since no LO decomposition can have more branches than the dimension of the smallest subsystem.  Conversely, any LO decomposition of $\ket{\psi}$ with a smaller number of branches can be fine-grained by repeatedly taking the Schmidt decomposition of the individual branches $\ket{\psi_i}$.

\section{Construction}

It is illuminating to explicitly construct the maximum decomposition in the case where all local density matrices are non-degenerate.  First diagonalize the local states,
\begin{align}
\rho^{(n)} =  \sum_\mu \lambda_\mu^{(n)} \ket{\mu}_n \bra{\mu},
\end{align}
with normalized eigenvectors $\ket{\mu}_n$.  
Then define
\begin{align} 
P_\mu^{(n)} = I^{\overline{(n)}} \otimes \ket{\mu}_n \bra{\mu}   
\end{align}
which projects onto $\ket{\mu}_n$ for the $n$-th subsystem and acts trivially otherwise.  
Now consider the graph (in the sense of graph theory) of all the eigenvectors of all $N$ subsystems. Let the $\mu$- and $\nu$-th eigenstates of two subsystems $n$ and $m$ be connected by an edge when  
\begin{align}
P_\mu^{(n)} P_\nu^{(m)} \ket{\psi} \neq 0.
\end{align}
(This is a symmetric definition since the two projectors commute.)  Then we can partition the graph into a unique set of connected subgraphs.  Index these subgraphs by $i$ and let $\mu \in T_i^{(n)}$ if and only if $\ket{\mu}_n$ is a part of the $i$-th connected subgraph.  Finally, for an arbitrary $n$, the terms in our maximal decomposition  are defined by   
\begin{align}
\sqrt{p_i}\ket{\psi_i} = Q_i^{(n)} \ket{\psi}
\end{align}
where
\begin{align}
Q_i^{(n)} = \sum_{\mu \in T_i^{(n)}} P_\mu^{(n)}
\end{align}
projects onto the eigenstates of subsystem $n$ in the $i$-th connected subgraph.  
To see that this definition does not depend on $n$, note that
\begin{align}\begin{split}
Q_i^{(n)} \ket{\psi} &= Q_i^{(n)} \left[Q_i^{(m)} + \sum_{j \neq i} Q_j^{(m)} \right] \ket{\psi} \\
&= Q_i^{(n)} Q_i^{(m)} \ket{\psi} \\
&= Q_i^{(m)} Q_i^{(n)} \ket{\psi} \\
&= Q_i^{(m)} \ket{\psi}
\end{split}\end{align}
where the projectors in the square brackets sum to unity on the support of $\ket{\psi}$ in $\Glob^{(m)}$. The second line follows from the definition of $T_i^{(n)}$, the third line from the fact that $Q_i^{(n)}$ and $Q_i^{(m)}$ commute, and the fourth line by repeating the first three lines with $n$ and $m$ exchanged.

To see that this is the maximal LO decomposition, consider a vector $\ket{\tilde{\psi}_k}$ in an arbitrary LO decomposition $\ket{\psi} = \sum_k \sqrt{\tilde{p}_k} \ket{\tilde{\psi}_k}$.  
Take any $\mu$, $i$, and $k$ such that $\mu \in \tilde{T}_k^{(n)}$ and $\mu \in T_i^{(n)}$.  We have
\begin{align}
P_\mu^{(n)} \ket{\tilde{\psi}_k} = P_\mu^{(n)} \ket{\psi} = P_\mu^{(n)} \ket{\psi_i}
\end{align}
because the projectors $P_\mu^{(n)}$ are defined independent of the decomposition. Then for any $\nu \in T_i^{(m)}$ we have
\begin{align}
P_\nu^{(m)} P_\mu^{(n)} \ket{\tilde{\psi}_k} = P_\nu^{(m)} P_\mu^{(n)} \ket{\psi} \neq 0.
\end{align}
Therefore, the $k$-conditional subspace of subsystem $m$ must contain $\ket{\nu}_m$, i.e.\ $\nu \in \tilde{T}_k^{(m)}$.  Iterating this process for all eigenvectors in the $i$-th connected subgraph shows that $T_i^{(m)} \subset \tilde{T}_k^{(m)}$ for all $m$.  In other words, the partitioning $\{ \tilde{T}_k^{(m)} \}_k$ is a coarse-graining of the partitioning $\{ T_i^{(m)} \}_i$ and the decomposition $\{ \ket{\tilde{\psi}_k} \}_k$ is a coarse-graining of the decomposition $\{ \ket{\psi_i} \}_i$ in the sense of \eqref{eq:coarse-graining}.

\section{Entanglement measure} 
\label{sec:entanglement}

Given a maximal LO decomposition $\ket{\psi} = \sum_i \sqrt{p_i} \ket{\psi_i}$, it's natural to define an entropy
\begin{align}
E_{\mathrm{LO}} \big[ \ket{\psi} \big] = S(\{ p_i \})
\end{align}
where $S(\{ p_i \})$ is (for instance) the Shannon entropy of the $p_i$'s.  For $N=2$ this reduces to the entropy of entanglement, just as the LO decomposition reduces to the Schmidt decomposition.

This entanglement measure is non-increasing under LOCC transformations.

There is a strong sense in which $E_{\mathrm{LO}} \big[ \ket{\psi} \big]$ measures strictly global entanglement in $\ket{\psi}$, i.e. entanglement involving all the subsystems together.  First, note that if any collection of the subsystems (including a single subsystem) is unentangled from the rest, then $E_{\mathrm{LO}} =0$.  Second, consider a LO decomposition that also happens to be constructed of product states.  It's most general form (in a properly chosen basis) is
\begin{align}
\label{eq:zstate}
\ket{Z} = \sum_i \sqrt{p_i} \ket{i}^{\otimes N}
\end{align}
This is a sort of GHZ state generalized to both many subsystems and to many unequally weighted branches. In this case, $E_{\mathrm{LO}} \big[ \ket{\psi} \big] = S(\{ p_i \})$ captures all the entanglement in the system.

More generally we can have entangling unitary operations $U_{n,m}$ which operate pairwise between subsystems $n$ and $m$.  So long as these unitaries do not take states $\ket{\mu}_n \ket{\nu}_m$ from one branch into the subspaces associated with the other branches, such operations leave the $p_i$ (and hence $E_{\mathrm{LO}}$) unchanged.  One example is the singlet-creating operation
\begin{align}
U_{n,m} =  \frac{1}{\sqrt{2}} \left(\begin{array}{rr}1 & 1 \\ 1 & -1 \end{array}\right)
\end{align}
acting on $\ket{Z}$, \eqref{eq:zstate}, in the basis $\{ \ket{i}_n \ket{i}_m, \ket{\tilde{i}}_n \ket{\tilde{i}}_m \}$,  where $\ket{\tilde{i}}_n$ denotes states outside the span of the $\ket{i}_n$ for each $n$.  Similar comments can be said about entangling operations between larger  proper subsets of the subsystems.

This entanglement measure is discussed further in the examples below.

\section{Examples}

For $N=3$ with $\Glob = \mcA \otimes \mcB \otimes \mcC$, consider the state
\begin{align}
\ket{U} \propto \Big(\ket{0}_\mcA \ket{0}_\mcB + \ket{1}_\mcA \ket{1}_\mcB \Big) \ket{0}_\mcC.
\end{align}
This state has no non-trivial LO decomposition, since the third subsystem is uncorrelated with the rest.  There is no information recorded everywhere, so it is natural that $E_{\mathrm{LO}} [\ket{U}] =0$.  Things do not change if we consider
\begin{align}\begin{split}
\ket{V} &\propto \Big(\ket{0}_\mcA\ket{0}_\mcB + \ket{1}_\mcA\ket{1}_\mcB \Big) \ket{0}_\mcC  \\
& \qquad  \qquad +\Big(\ket{0}_\mcA\ket{2}_\mcB + \ket{1}_\mcA\ket{3}_\mcB \Big) \ket{1}_\mcC \\
&= \ket{0}_\mcA \Big(\ket{0}_\mcB\ket{0}_\mcC + \ket{2}_\mcB\ket{1}_\mcC \Big)  \\
& \qquad \qquad  + \ket{1}_\mcA \Big(\ket{1}_\mcB\ket{0}_\mcC + \ket{3}_\mcB\ket{1}_\mcC \Big),
\end{split}\end{align}
where the first and second subsystems share an entangled qubit, as do the second and third. This means that all local density matricies are mixed.  But still, there is no non-trivial decomposition because there is still no common piece of information that is recorded everywhere.  This continues to hold if we consider the state $\ket{X}$ (studied previously \cite{bennett2000exact}) constructed by having three entangled bits shared pairwise symmetrically between the three subsystems.

The well-known $W$ state does not have a non-trivial LO decomposition,
\begin{align}
\ket{W} \propto \ket{001} + \ket{010} +\ket{100},
\end{align}
whereas the GHZ state does manifestly,
\begin{align}
\ket{\mathrm{GHZ}} \propto \ket{000} + \ket{111}.
\end{align}

\begin{acknowledgments}
I thank M\=aris Ozols, Graeme Smith, John Smolin, Wojciech Zurek, and Michael Zwolak for discussion.  This research was supported by the John Templeton Foundation through grant number 21484.
\end{acknowledgments}

\bibliographystyle{apsrev4-1}
\bibliography{riedelbib}
\end{document}

%% file: riedelheader.tex
\usepackage{
	amssymb, 
	amsthm, 
	graphicx, 
	xspace,
	overpic,
	ifthen,
	verbatim, 
	subfigure, 
	appendix,
	ifthen,
	nicefrac, 
	color, 
	tensor, 
	etoolbox, 
	centernot, 
	MnSymbol, 
	amsmath,
	bm
} 

\newcommand{\bra}[1]{\langle #1 \rvert  }
\newcommand{\ket}[1]{\lvert #1 \rangle  }

\newcommand{\braket}[2]{  \langle #1 \vert #2 \rangle  }

\newcommand{\Glob}{\ensuremath{\mathcal{H}}}

\newcommand{\mcA}{\ensuremath{\mathcal{A}}}
\newcommand{\mcB}{\ensuremath{\mathcal{B}}}
\newcommand{\mcC}{\ensuremath{\mathcal{C}}}

\newcommand{\refstyle}{2}  
\newcommand{\eqrefstyle}{2}  
\newcommand{\Jeqref}[1] {\ifnum\refstyle=1{Eq.~(\ref{#1})}\else{equation (\ref{#1})}\fi}
\newcommand{\Jeqsref}[1] {\ifnum\refstyle=1{Eqs.~(\ref{#1})}\else{equations (\ref{#1})}\fi}
\newcommand{\EQref}[1] {\ifnum\eqrefstyle=1{(\ref{#1})}\else{\Jeqref{#1}}\fi}
\newcommand{\EQsref}[1] {\ifnum\eqrefstyle=1{(\ref{#1})}\else{\Jeqsref{#1}}\fi}
\newcommand{\Eqref}[1]  {\ifnum\eqrefstyle=3{\Jeqref{#1}}\else{(\ref{#1})}\fi}
\newcommand{\Eqsref}[1]  {\ifnum\eqrefstyle=3{\Jeqsref{#1}}\else{(\ref{#1})}\fi}
\newcommand{\Figref}[1]{\ifnum\refstyle=1{Fig.~\ref{#1}}\else{figure~\ref{#1}}\fi}
\newcommand{\Figsref}[1]{\ifnum\refstyle=1{Figs.~\ref{#1}}\else{figures~\ref{#1}}\fi}
\newcommand{\Appref}[1]{\ifnum\refstyle=1{App.~\ref{#1}}\else{appendix~\ref{#1}}\fi}
\newcommand{\Appsref}[1]{\ifnum\refstyle=1{Apps.~\ref{#1}}\else{appendices~\ref{#1}}\fi}
\newcommand{\Chref}[1]{\ifnum\refstyle=1{Ch.~\ref{#1}}\else{chapter~\ref{#1}}\fi}
\newcommand{\Chsref}[1]{\ifnum\refstyle=1{Chs.~\ref{#1}}\else{chapters~\ref{#1}}\fi}
\newcommand{\Secref}[1]{\ifnum\refstyle=1{Sec.~\ref{#1}}\else{section~\ref{#1}}\fi}
\newcommand{\Secsref}[1]{\ifnum\refstyle=1{Secs.~\ref{#1}}\else{sections~\ref{#1}}\fi}
\newcommand{\Refref}[1]{\ifnum\refstyle=1{Ref.~\cite{#1}}\else{reference~\cite{#1}}\fi}
\newcommand{\Refsref}[1]{\ifnum\refstyle=1{Refs.~\cite{#1}}\else{references~\cite{#1}}\fi}

\newcommand{\draftmode}{1}    
\newcommand{\notetoself}[1]{\ifnum \draftmode=1 {\color[rgb]{0,0,0.8} [#1]} \fi}  
\newcommand{\cuttext}[1]{\ifnum \draftmode=1 {\color[rgb]{0,0.5,0} [#1]} \fi}  
\newcommand{\warntext}[1]{\ifnum \draftmode=1 {\color[rgb]{0.9,0.6,0} #1} \else {#1} \color{black} \fi}  